\documentclass[12pt]{article}
\newcommand{\be}{\begin{equation}}
\newcommand{\ee}{\end{equation}}

\newcommand{\bi}{\begin{itemize}}
\newcommand{\ei}{\end{itemize}}
\newcommand{\bea}{\begin{eqnarray}}
\newcommand{\eea}{\end{eqnarray}}
\newcommand{\ba}{\begin{array}}
\newcommand{\ea}{\end{array}}

\usepackage[left=2.50cm, right=2.50cm, top=2.50cm, bottom=2.50cm]{geometry}
\usepackage[utf8]{inputenc}
\usepackage{cancel}
\usepackage[english]{babel}
\usepackage{physics}
\usepackage{hyperref}
\usepackage{amsthm}
\usepackage{mathrsfs}
\usepackage{mathtools}
\usepackage{graphicx}
\usepackage{changepage}
\usepackage{amssymb,amsmath}
\usepackage{verbatim}
\usepackage{empheq}
\usepackage{xcolor}
\usepackage{tikz}
\usepackage{float}
\usepackage{multirow}
\usepackage{multicol}
\usepackage{amsmath}
\usepackage{centernot}
\usepackage[hang,flushmargin]{footmisc} 
\usepackage[normalem]{ulem}
\usepackage[noadjust]{cite}
\usetikzlibrary{tikzmark}
\numberwithin{equation}{section}
\hypersetup{hidelinks}

\newlength{\bibitemsep}
\newlength{\bibparskip}
\let\oldthebibliography\thebibliography
\renewcommand\thebibliography[1]{%
  \oldthebibliography{#1}%
  \setlength{\parskip}{\bibitemsep}%
  \setlength{\itemsep}{\bibparskip}%
}
\begin{document}
\par
\bigskip
\Large
\noindent
{\bf 
Emergent fracton strings from covariant bi-form gauge field theory}
\bigskip
\par
\rm
\normalsize

\hrule

\vspace{1cm}

\large
\noindent
{\bf Erica Bertolini$^{1,a}$},
{\bf Hyungrok Kim$^{2,b}$}, 
{\bf Giandomenico Palumbo$^{3,c}$}\\

\par

\small

\noindent$^1$ School of Theoretical Physics, Dublin Institute for Advanced Studies, 10 Burlington Road, D04 C932, Dublin, Ireland.

\noindent$^2$ Centre for Mathematics and Theoretical Physics Research, Department of Physics, Astronomy and Mathematics, University of Hertfordshire, Hatfield, Hertfordshire AL10 9AB, United Kingdom.

\noindent$^3$ CFisUC, Department of Physics, University of Coimbra, Rua Larga, 3004-516 Coimbra, Portugal.

\smallskip

\smallskip

\vspace{1cm}

\noindent
{\tt Abstract}\\
We present a covariant field-theoretical framework for a rank-4 tensor gauge field theory describing fractonic string-like objects. We show that the most general quadratic, parity-preserving action naturally leads to a Maxwell-like sector, with tensorial analogues of electric and magnetic fields, Maxwell-like equations, energy-momentum tensor, and a Lorentz-like force. Remarkably, the theory gives rise to fracton-like string excitations purely from symmetry principles: constraints on the motion of these extended objects appear as Gauss-like laws, without being imposed by hand. One of these laws is new and corresponds to a generalised dipole conservation for closed strings, restricting their mobility and defining a novel class of fractonic string-like excitations.  Finally, we uncover a connection to linearised area-metric gravity: in a suitable limit, the theory reduces to known covariant fracton models with rank-2 gauge fields, highlighting a deep link between fractonic matter and gravity-like structures. This provides a unified perspective on higher-rank gauge fields, extended excitations, and emergent gravitational features.
\\

\vspace{\fill}

\noindent{\tt Keywords:} Quantum field theory, fractons, Maxwell gauge field theory, tensor gauge field theory, strings, area metric gravity \\

\vspace{1cm}

\hrule
\noindent{\tt E-mail:
$^a$ebertolini@stp.dias.ie,
$^b$h.kim2@herts.ac.uk,
$^c$giandomenico.palumbo@gmail.com.
}
\newpage
\tableofcontents
\newpage
\section{Introduction}

Fracton phases \cite{Vijay:2016phm,Nandkishore:2018sel,Pretko:2020cko,Gromov:2022cxa} emerged as a distinctive arena of condensed-matter physics in which excitations exhibit sharply restricted mobility, associated to generalised conservation laws of multipole momenta \cite{Gromov:2018nbv,Caddeo:2022ibe,Bidussi:2021nmp,Hartong:2024hvs}, and gauge structures beyond ordinary vector electromagnetism \cite{Pretko:2016lgv}. In higher-rank $U(1)$ tensor gauge theories \cite{Pretko:2016kxt}, these features are evident. For example in the rank-2 ``scalar charge theory of fractons'' \cite{Pretko:2016lgv,Pretko:2016kxt} the Gauss-like law
	\be
	\partial_i\partial_jE^{ij}=\rho\ ,
	\ee
is considered, for a symmetric electric tensor $E^{ij}(x)$ being the conjugate momentum for the theory. That constraint, together with its associated continuity equation 
	\be
	\partial_t\rho+\partial_i\partial_jJ^{ij}=0\ ,
	\ee
lie at the heart of dipole conservation \cite{Pretko:2016lgv,Gromov:2018nbv}, in the sense that the total dipole of the theory, $D^i$, must not change
	\be\label{dipole cons}
	\partial_tD^i=\int dV x^i\partial_t\rho=0\ .
	\ee
That thus implies the immobility of the fractonic charge density $\rho(x)$, while its dipole density is free to move \cite{Pretko:2016lgv}. Similarly, other rank-2 models are built by adding constraints to the electric tensor (such as tracelessness), and/or by modifying the nature of the Gauss law, which typically changes the nature of the fractonic charge itself (from scalar to vector). These introduce new constraints and conservation laws that produce new kinds of quasiparticles with different kinds of mobilities, such as the so-called lineons and planons \cite{Pretko:2020cko,Seiberg:2020wsg,Gorantla:2022pii,Figueroa-OFarrill:2025yum}. From a quantum field theoretical perspective, these systems invite a formulation in which symmetry, covariance, locality, and power-counting are the building blocks that systematise the space of allowed theories as consequences of first principles rather than \textit{ad hoc} conditions \cite{Bertolini:2022ijb,Bertolini:2023sqa,Blasi:2022mbl,Bertolini:2023juh,Bertolini:2024apg,Bertolini:2025qcy,Bertolini:2025jov,Bertolini:2025jul,Bertolini:2025goo,Afxonidis:2023pdq,Rovere:2024nwc,Rovere:2025nfj,Fecit:2025eet,Liang:2025gpp,Hinterbichler:2025ost}. These descriptions are particularly valuable when one seeks unifying principles, controlled couplings to matter currents, and connections to geometric frameworks that have proven fruitful across high-energy and condensed-matter contexts. Strings and other higher-dimensional defects are a generic feature of quantum field theories, with close connections to (generalised) symmetries \cite{Gaiotto:2014kfa}. Strings and branes in the tensionless limit may be considered simply as a collection of massless point particles subject to certain constraints \cite{Lindstrom:2022iec,Borsten:2025dcx}, thus exhibiting fractonic behaviour; tensile strings and branes similarly exhibit an infinite tower of fractonic constraints \cite{Bertolini:2024nmn}. Thus, ordinary strings and branes fit into a spectrum of fractonic constraints that includes ordinary particles, fractons, lineons and planons as well. In between ordinary strings and point-like fractons are strings with further fractonic constraints, which are known to arise in nonrelativistic systems \cite{Pai:2018qnm}. Since relativistic systems are known to harbour both point-like fractons \cite{Bertolini:2022ijb} as well as strings and branes, it is a natural question to see if fractonic strings can also arise there. A first hint towards a positive answer was found in \cite{Makino:2025mzo}. Besides string theory, a broader motivation for investigating fractonic behaviours beyond the point-like regime comes from the fact that condensed matter naturally hosts not only point-like quasiparticles, but also extended objects. For instance, in four spacetime dimensions, superconducting vortices \cite{Horn:2015zna, Pretko:2017khx}  and disclination lines \cite{Schimming} behave as string-like excitations, but extended objects also appear in interacting topological phases \cite{Levin:2004mi,Palumbo:2024hym,Palumbo:2026nua}, skyrmion systems \cite{yokouchi2018,Palumbo:2021dkx,Gudnason:2025inp}, and spin liquids \cite{Chung:2023mqw}. It is then natural to ask whether fractonic behaviour can occur for such extended objects, namely, whether there exist \emph{string-like} excitations whose motion is constrained in a genuinely fractonic way. A key step in this direction was taken by Pai and Pretko \cite{Pai:2018qnm}, who exhibited ``fractonic line'' excitations and formulated a noncovariant rank-4 tensor gauge theory description tied to elasticity and its lattice defects by means of the known duality between the two systems \cite{Pretko:2017kvd,Gromov:2017vir,Pretko:2019omh,Gromov:2019waa,Grosvenor:2021hkn,Tsaloukidis:2023jmr,Tsaloukidis:2023bvz,Nguyen:2020yve}. Similarly, in \cite{Shenoy:2019wng}, a noncovariant rank-$k$ theory for extended fractons were studied. In the context of fractonic lattice gauge theories \cite{Vijay:2016phm,Haah:2011drr,Vijay:2015mka}, fractonic-line excitations of that kind were also studied in \cite{Li:2019tje,Qi:2020jrf}. Despite this progress, a fully covariant field theory of fracton strings, $i.e.$ one that produces the relevant mobility restrictions from symmetry principles and admits the familiar structural hallmarks of gauge theory, is still missing. Therefore it is the aim of this work to address this gap by constructing a new gauge field theory for a rank-4 (bi-form) tensor gauge field \cite{deMedeiros:2002qpr,deMedeiros:2003qel,Hinterbichler:2022agn} in four spacetime dimensions within a covariant formalism and symmetry principles, inspired by the noncovariant model of \cite{Pai:2018qnm}, and the successful covariant theory of rank-2 fractons \cite{Bertolini:2022ijb}.\\

The paper is organised as follows. In Section \ref{section 2} we build the most general invariant theory. In particular  in Section \ref{ingredients} we introduce the gauge field, its symmetry transformation, and invariant field strength. In Section \ref{section-general eom} the invariant action is defined compatibly with covariance, locality, power-counting and parity, its equations of motion are computed together with the conjugate momentum. A Gauss-like law for the full theory is also identified. In Section \ref{section-area-metric} we remark the relations with area-metric gravity and the rank-2 covariant fracton theory of \cite{Bertolini:2022ijb}. Section \ref{sec3} is devoted to the Maxwell-like fractonic theory. Thus after tuning the action, in Section \ref{section-em-like} the full electromagnetic analogy is exploited with its electric and magnetic tensor fields, while in Section \ref{section-sources} extended  fractonic matter sources are introduced in the action and the charges are identified as closed strings and their dipoles, each sourcing a Gauss law. These laws are then analysed in Section \ref{section-constraints}, and the conserved quantities are found to fully constrain the strings, which are thus fractonic. Finally in Section \ref{section-lorentz} the energy-momentum tensor is computed and a Lorentz-like force for the dipole of the string is recovered. In Section \ref{sec-conclusion} we comment and summarise our results.\\

{\bf Notations}

$4D=3+1$ spacetime dimensions .\\
Indices:  $\mu,\nu,\rho,...=\{0,1,2,3\}\ \,\ i,j,k,...=\{1,2,3\}$ .\\
Minkowski metric: $ \eta_{\mu\nu}=\mbox{diag}(-1,1,1,1)\ . $ \\
Levi-Civita symbol: $\epsilon^{0123}=1=-\epsilon_{0123}\ .$

\section{Extended fractons and area-metric gravity}\label{section 2}

\subsection{Gauge field, symmetry and field strength}\label{ingredients}
We consider the rank-4 (bi-form) gauge field $A_{\mu\nu|\rho\sigma}(x)$ with the following properties on the indices
	\be
	A_{\mu\nu|\rho\sigma}=A_{\rho\sigma|\mu\nu}\quad;\quad A_{\mu\nu|\rho\sigma}=-A_{\nu\mu|\rho\sigma}=-A_{\mu\nu|\sigma\rho}\ ,
	\ee
and
	\be
	A_{\mu\nu|\rho\sigma}+A_{\mu\rho|\sigma\nu}+A_{\mu\sigma|\nu\rho}=0\ .
	\ee
The gauge field transforms under the following gauge transformation
	\be\label{symm}
	\delta A_{\mu\nu|\rho\sigma}=\partial_\nu\partial_\rho\lambda_{\mu\sigma}-\partial_\mu\partial_\rho\lambda_{\nu\sigma}+\partial_\mu\partial_\sigma\lambda_{\nu\rho}-\partial_\nu\partial_\sigma\lambda_{\mu\rho}\ ,
	\ee
with a symmetric rank-2 gauge parameter $\lambda_{\mu\nu}=\lambda_{\nu\mu}$. This transformation is the covariant generalisation of the one introduced in certain noncovariant rank-4 fracton models \cite{Pai:2018qnm}, which therefore are embedded in our theory. Bi-form gauge fields arise in the general framework of mixed-symmetry tensor gauge theories, where the gauge potential carries two independent antisymmetric sets of indices and can be interpreted geometrically as a differential form valued in another exterior algebra. Such structures naturally generalize ordinary $p$-form gauge theories and have been studied in several contexts, including generalized gauge fields and dual formulations of higher-rank tensor theories \cite{deMedeiros:2002qpr,deMedeiros:2003qel,Hinterbichler:2022agn}. As we will see, this geometric framework, together with its gauge transformation \eqref{symm}, will provide a natural covariant setting to encode the constrained dynamics and mobility restrictions characteristic of fractonic string excitations. Notice that there is a freedom on the gauge parameter $\lambda_{\mu\nu}(x)$ which is a kind of diffeomorphism invariance
	\be
	\lambda_{\mu\nu}\to\bar\lambda_{\mu\nu}=\lambda_{\mu\nu}+\partial_\mu\xi_\nu+\partial_\nu\xi_\mu\ ,
	\ee
due to which a gauge choice for $\lambda_{\mu\nu}(x)$ needs to be considered as well. For instance a possible choice is $\lambda_{\mu0}(x)=0$ which will be later taken into account for the physical interpretation of the theory. We define the traces of the gauge field $A_{\mu\nu|\rho\sigma}(x)$ as
	\be
	A_{\mu\nu}=A_{\nu\mu}\equiv \eta^{\alpha\beta}A_{\mu\alpha|\nu\beta}=A_{\mu\ |\nu\lambda}^{\ \lambda}=A^\lambda_{\ \mu|\lambda\nu}\qquad;\qquad A\equiv\eta^{\mu\nu}\eta^{\alpha\beta}A_{\mu\alpha|\nu\beta}= A^{\mu\nu|}_{\quad\mu\nu}\ ,
	\ee
whose transformations are
	\be
	\delta A_{\mu\nu}=\partial_\mu\partial^\alpha\lambda_{\nu\alpha}+\partial_\nu\partial^\alpha\lambda_{\mu\alpha}-\Box\lambda_{\mu\nu}-\partial_\mu\partial_\nu\lambda\qquad;\qquad\delta A=2\left(\partial^\mu\partial^\nu\lambda_{\mu\nu}-\Box\lambda\right)\ ,
	\ee
with $\lambda(x)\equiv\eta^{\mu\nu}\lambda_{\mu\nu}(x)$. A rank-5 invariant field strength, \emph{i.e.}\ such that $\delta F^{\alpha\mu\nu|\rho\sigma}(x)=0$, can be defined as follows:
	\be
	F_{\alpha\mu\nu|\rho\sigma}\equiv 2\partial_\alpha A_{\mu\nu|\rho\sigma}-\partial_\mu A_{\rho\sigma|\alpha\nu}+\partial_\nu A_{\rho\sigma|\alpha\mu}-\partial_\rho A_{\mu\nu|\alpha\sigma}+\partial_\sigma A_{\mu\nu|\alpha\rho}\ ,\label{F}
	\ee
whose contractions are
	\begin{align}
	F_{\alpha\mu\nu}=F_{\alpha\nu\mu}\equiv\,&\eta^{\beta\gamma}F_{\alpha\beta\mu|\gamma\nu}=-\partial^\beta\left(A_{\beta\nu|\alpha\mu}+A_{\beta\mu|\alpha\nu}\right)+2\partial_\alpha A_{\mu\nu}-\partial_\mu A_{\alpha\nu}-\partial_\nu A_{\alpha\mu}\label{Fmunu}\\
&\eta^{\alpha\beta}F_{\alpha\beta\mu|\nu\rho}=-\tfrac{1}{3}\left(F_{\nu\mu\rho}-F_{\rho\nu\mu}\right)\\
	F_\alpha\equiv&\,\eta^{\mu\nu}F_{\alpha\mu\nu}=-2\eta^{\mu\nu}F_{\mu\nu\alpha}=2\left(\partial_\alpha A-2\partial^\mu A_{\mu\alpha}\right)\label{Fmu}\ ,
	\end{align}
which represent the building blocks in terms of which all the physical quantities will be expressed in order to make them invariant at sight. This field strength is characterised by the following index properties
	\begin{align}
	F_{\alpha\mu\nu|\rho\sigma}=F_{\alpha\rho\sigma|\mu\nu}\qquad;\qquad &F_{\alpha\mu\nu|\rho\sigma}=-F_{\alpha\nu\mu|\rho\sigma}=-F_{\alpha\mu\nu|\sigma\rho}\\
	F_{\alpha\mu\nu|\rho\sigma}+F_{\alpha\mu\rho|\sigma\nu}+F_{\alpha\mu\sigma|\nu\rho}=0\qquad;\qquad &F_{\alpha\mu\nu}+F_{\mu\nu\alpha}+F_{\nu\alpha\mu}=0\label{ciclicity}\\
	F_{\alpha\mu\nu|\rho\sigma}+\tfrac1 2\left(F_{\mu\alpha\nu|\rho\sigma}+F_{\nu\mu\alpha|\rho\sigma}\!\right.&\left.+F_{\rho\mu\nu|\alpha\sigma}+F_{\sigma\mu\nu|\rho\alpha}\right)=0\ ,\label{cycF2}
	\end{align}
and it satisfies the Bianchi-like identity
	\be
	\epsilon^{\mu\nu\rho\sigma}\partial_\mu F_{\nu\alpha\beta|\rho\sigma}=\epsilon^{\mu\nu\rho\sigma}\partial_\mu F_{\alpha\beta\nu|\rho\sigma}=0\ .\label{bianchi}
	\ee

\subsection{Action, equations of motion and Gauss-like law}\label{section-general eom}

The only independent,local, covariant, parity-preserving, quadratic terms in the action are, up to boundary terms, the following ones
	\begin{align}
	A^2\quad;\quad A_{\mu\nu}A^{\mu\nu}\quad&;\quad A_{\mu\nu|\rho\sigma}A^{\mu\nu|\rho\sigma}\\
	A\Box A\quad;\quad A_{\mu\nu}\Box A^{\mu\nu}\quad&;\quad A_{\mu\nu|\rho\sigma}\Box A^{\mu\nu|\rho\sigma} \\
	A\partial_\mu\partial_\nu A^{\mu\nu}\quad;\quad A_{\mu\nu}\partial^\nu\partial_\rho A^{\mu\rho}\quad&;\quad A_{\mu\nu}\partial_\rho\partial_\sigma A^{\mu\rho|\nu\sigma}\quad;\quad A_{\mu\nu|\rho\sigma}\partial^\mu\partial_\lambda A^{\lambda\nu|\rho\sigma}\ ,
	\end{align}
for which the most general action invariant under the gauge symmetry \eqref{symm} is found to be
	\be
		\begin{split}
		S_{inv}&=\int d^4x\bigl[a_1\left(A\Box A-4A\partial_\mu\partial_\nu A^{\mu\nu}+4A_{\mu\nu}\partial^\nu\partial_\rho A^{\mu\rho}\right)\\
		&\qquad\qquad +a_2\left( A_{\mu\nu}\Box A^{\mu\nu}-2A_{\mu\nu}\partial_\rho\partial_\sigma A^{\mu\rho|\nu\sigma}-A_{\mu\nu}\partial^\nu\partial_\rho A^{\mu\rho}+\tfrac{1}{2}A_{\mu\nu|\rho\sigma}\partial^\mu\partial_\alpha A^{\alpha\nu|\rho\sigma}\right)\\
		&\qquad\qquad\left.\! +a_3\left(A_{\mu\nu|\rho\sigma}\Box A^{\mu\nu|\rho\sigma}-2A_{\mu\nu|\rho\sigma}\partial^\mu\partial_\alpha A^{\alpha\nu|\rho\sigma}\right)\right]\\
		&=\int d^4x\left(-\tfrac{a_1}{4}F_\alpha F^\alpha-\tfrac{a_2}{6}F_{\alpha\mu\nu}F^{\alpha\mu\nu}-\tfrac{a_3}{8}F_{\alpha\mu\nu|\rho\sigma}F^{\alpha\mu\nu|\rho\sigma}\right)\ ,\label{Sinv}
		\end{split}
	\ee
with $a_1,\ a_2,\ a_3$ free dimensionless coefficients, being the mass dimension of the gauge field $[A]=1$. One of these coefficients can be reabsorbed by a field redefinition, however, we will keep all of them explicit in order to better analyse any particular combination and, eventually, tune them. Notice that no massive terms are allowed by the symmetry. Other possible invariant terms exist, such as
	\be
	\epsilon^{\mu\nu\rho\sigma}F_{\alpha\beta\gamma|\mu\nu}F^{\alpha\beta\gamma|}_{\ \ \quad\rho\sigma}\quad;\quad\epsilon^{\mu\nu\rho\sigma}F_{\mu\nu\alpha}F_{\rho\sigma}^{\ \ \alpha}
	\ee
that remind the theta-term of Yang-Mills theory \cite{Coleman:1985rnk}, but without necessarily being pure boundary contributions. However these are parity-violating terms, and will not be considered for the present work. Keeping in mind that
	\begin{align}
		\begin{split}
		\frac{\delta A_{\mu\nu|\rho\sigma}}{\delta A_{\alpha\beta|\gamma\lambda}}&=\frac1 2\left(\delta^\alpha_\mu\delta^\beta_\nu\delta^\gamma_\rho\delta^\lambda_\sigma-\delta^\alpha_\nu\delta^\beta_\mu\delta^\gamma_\rho\delta^\lambda_\sigma+\delta^\alpha_\nu\delta^\beta_\mu\delta^\gamma_\sigma\delta^\lambda_\rho-\delta^\alpha_\mu\delta^\beta_\nu\delta^\gamma_\sigma\delta^\lambda_\rho\right.\\
		&\quad\qquad\left.+\delta^\alpha_\rho\delta^\beta_\sigma\delta^\gamma_\mu\delta^\lambda_\nu-\delta^\alpha_\sigma\delta^\beta_\rho\delta^\gamma_\mu\delta^\lambda_\nu+\delta^\alpha_\sigma\delta^\beta_\rho\delta^\gamma_\nu\delta^\lambda_\mu-\delta^\alpha_\rho\delta^\beta_\sigma\delta^\gamma_\nu\delta^\lambda_\mu\right)
		\end{split}\\
		\begin{split}
		\frac{\delta A_{\mu\nu}}{\delta A_{\alpha\beta|\gamma\lambda}}&=\frac1 2\left[-\eta^{\alpha\lambda}\left(\delta^\beta_\mu\delta^\gamma_\nu+\delta^\beta_\nu\delta^\gamma_\mu\right)+\eta^{\alpha\gamma}\left(\delta^\beta_\mu\delta^\lambda_\nu+\delta^\beta_\nu\delta^\lambda_\mu\right)\right.\\
		&\qquad\ \left.-\eta^{\beta\gamma}\left(\delta^\alpha_\mu\delta^\lambda_\nu+\delta^\alpha_\nu\delta^\lambda_\mu\right)+\eta^{\beta\lambda}\left(\delta^\alpha_\mu\delta^\gamma_\nu+\delta^\alpha_\nu\delta^\gamma_\mu\right)\right]
		\end{split}\\
		\begin{split}
		\frac{\delta A}{\delta A_{\alpha\beta|\gamma\lambda}}&=2\left(\eta^{\alpha\gamma}\eta^{\beta\lambda}-\eta^{\beta\gamma}\eta^{\alpha\lambda}\right)\ ,
		\end{split}
	\end{align}
the equations of motion (EoM) of the theory described by the invariant action $S_{inv}$ \eqref{Sinv} are
	\be\label{EoMgen}
		\begin{split}
		\frac{\delta S_{inv}}{\delta A_{\mu\nu|\rho\sigma}}=&a_1\left[2\left(\eta^{\mu\rho}\eta^{\nu\sigma}-\eta^{\nu\rho}\eta^{\mu\sigma}\right)\partial_\alpha F^\alpha+  \eta^{\mu\sigma}\left(\partial^\nu F^\rho+\partial^\rho F^\nu\right)-  \eta^{\mu\rho}\left(\partial^\nu F^\sigma+\partial^\sigma F^\nu\right)\right.\\
		&\quad\left.+  \eta^{\nu\rho}\left(\partial^\mu F^\sigma+\partial^\sigma F^\mu\right)-  \eta^{\nu\sigma}\left(\partial^\mu F^\rho+\partial^\rho F^\mu\right)\right]\\
		&+a_2\left[\partial_\alpha\left(\eta^{\mu\rho}F^{\alpha\nu\sigma}+\eta^{\nu\sigma}F^{\alpha\mu\rho}-\eta^{\mu\sigma}F^{\alpha\nu\rho}-\eta^{\nu\rho}F^{\alpha\mu\sigma}\right)+\tfrac1 3\partial^\mu\left(F^{\sigma\nu\rho}-F^{\rho\nu\sigma}\right)\right.\\
		&\quad\left.+\tfrac1 3\partial^\nu\left(F^{\rho\mu\sigma}-F^{\sigma\mu\rho}\right)+\tfrac1 3\partial^\rho\left(F^{\nu\sigma\mu}-F^{\mu\sigma\nu}\right)+\tfrac1 3\partial^\sigma\left(F^{\mu\nu\rho}-F^{\nu\mu\rho}\right)\right]\\
		&+4a_3\partial_\alpha F^{\alpha\mu\nu|\rho\sigma}\ ,
		\end{split}
	\ee
whose traces are
	\begin{align}
	\frac{\delta S_{inv}}{\delta A_{\mu\nu}}=&\left(4a_1+a_2\right)\left[\eta^{\mu\nu}\partial_\alpha F^\alpha-\tfrac1 2\left(\partial^\mu F^\nu+\partial^\nu F^\mu\right)\right]+\left(a_2+4a_3\right)\partial_\alpha F^{\alpha\mu\nu}\label{eomAmunu}\\
	\frac{\delta S_{inv}}{\delta A}=&2\left(6a_1+2a_2+2a_3\right)\partial_\alpha F^\alpha\ ,\label{eomA}
	\end{align}
where we used the fact that $F_\mu^{\ \mu\nu}=-\tfrac1 2 F^\nu$ \eqref{Fmu}. The conjugate momentum of the theory is given by
	\be\label{Pi-tot}
		\begin{split}
		\Pi^{\mu\nu|\rho\sigma}&=\frac{\delta S_{inv}}{\delta\partial_t A_{\mu\nu|\rho\sigma}}\\
		&=-a_1\left[2\left(\eta^{\mu\rho}\eta^{\nu\sigma}-\eta^{\mu\sigma}\eta^{\nu\rho}\right)F^0+\left(\eta^{\nu\rho}\eta^{\sigma0}-\eta^{\nu\sigma}\eta^{\rho0}\right)F^\mu+\left(\eta^{\mu\sigma}\eta^{\rho0}-\eta^{\mu\rho}\eta^{\sigma0}\right)F^\nu+\right.\\
		&\qquad\left.+\left(\eta^{\mu\sigma}\eta^{\nu0}-\eta^{\nu\sigma}\eta^{\mu0}\right)F^\rho+\left(\eta^{\nu\rho}\eta^{\mu0}-\eta^{\mu\rho}\eta^{\nu0}\right)F^\sigma\right]+\\
		&\quad+a_2\left\{\tfrac1 3 \left[\eta^{\mu0}\left(F^{\rho\sigma\nu}-F^{\sigma\rho\nu}\right)+\eta^{\nu0}\left(F^{\sigma\rho\mu}-F^{\rho\sigma\mu}\right)+\eta^{\rho0}\left(F^{\mu\nu\sigma}-F^{\nu\mu\sigma}\right)+\right.\right.\\
		&\qquad\left.+\left.\eta^{\sigma0}\left(F^{\nu\mu\rho}-F^{\mu\nu\rho}\right)\right]-\left(\eta^{\mu\rho}F^{0\nu\sigma}+\eta^{\nu\sigma}F^{0\mu\rho}-\eta^{\mu\sigma}F^{0\nu\rho}-\eta^{\nu\rho}F^{0\mu\sigma}\right)\right\}-\\
		&\quad-4a_3F^{0\mu\nu|\rho\sigma}\ ,
		\end{split}
	\ee
with traces
	\begin{align}
	\Pi^{\mu\nu}&\equiv\Pi^{\mu\lambda|\nu}_{\ \,\quad\lambda}=\left(4a_1+a_2\right)\left[\tfrac1 2 \left(\eta^{\mu0}F^\nu+\eta^{\nu0}F^\mu\right)-\eta^{\mu\rho}F^0\right]-\left(a_2+4a_3\right)F^{0\mu\nu}\,\\
	\Pi&\equiv\Pi^{\mu}_{\ \mu}=-2\left(6a_1+2a_2+2a_3\right)F^0\ .
	\end{align}
The conjugate momentum $\Pi^{\mu\nu|\rho\sigma}(x)$ \eqref{Pi-tot} is one of the key ingredients that are necessary to identify a fractonic theory. That is because the conjugate momentum, in those cases, is associated to an electric tensor field which, on its turn, is required to satisfy the Gauss-like law which is the key to obtain the limited mobility property of the fracton quasiparticles \cite{Pretko:2016kxt}. Notice that
	\begin{align}
	\Pi^{\mu\nu}=0\qquad &\Leftrightarrow\qquad a_2=-4a_1\quad ;\quad a_3=a_1\label{traceless1}\\
	\Pi=0\qquad &\Leftrightarrow\qquad a_3=-(3a_1+a_2)\ ,\label{traceless2}
	\end{align}
which signal that in these cases the theory does not depend on the trace $A_{\mu\nu}(x)$ and/or $A(x)$. Additionally we have that
	\be
	\Pi^{00|00}=\Pi^{m0|00}=\Pi^{m0|n0}=0\ ,
	\ee
and therefore the only nonvanishing, independent components are
	\begin{align}
	\Pi^{mn|pq}&=-2a_1\left(\eta^{mp}\eta^{nq}-\eta^{mq}\eta^{np}\right)F^0-a_2\left(\eta^{mp}F^{0nq}+\eta^{nq}F^{0mp}-\eta^{mq}F^{0np}-\eta^{np}F^{0mq}\right)-4a_3F^{0mn|pq}\\
	\Pi^{mn|p0}&=a_1\left(\eta^{np}F^m-\eta^{mp}F^n\right)+a_2\left[\tfrac1 3 \left(F^{mnp}-F^{nmp}\right)-\eta^{mp}F^{00n}+\eta^{np}F^{00m}\right]-4a_3F^{0mn|p0}\ .
	\end{align}
Having analysed the properties of the conjugate momentum we now turn into analysing the components of the EoM \eqref{EoMgen} in order to see whether, in analogy to the rank-2 covariant fracton case \cite{Bertolini:2022ijb}, a Gauss-like law already appears at the level of the full theory.
	\bi
	\item Considering $\mu=m,\ \nu=0,\ \rho=n,\ \sigma=0$ in the EoM \eqref{EoMgen} we get
		\be\label{EoMm0n0}
			\begin{split}
			\frac{\delta S_{inv}}{\delta A_{m0|n0}}&=a_1\left(-2\eta^{mn}\partial_aF^a+\partial^mF^n+\partial^nF^n\right)+\\
		 	&\quad+a_2\left(\eta^{mn}\partial_aF^{a00}-\partial_aF^{amn}-\tfrac1 2 \partial^mF^{n00}-\tfrac1 2 \partial^mF^{n00}\right)+4a_3\partial_aF^{am0|n0}\\
			&=-\partial_a\left(\Pi^{am|n0}+\Pi^{an|m0}\right)\ .
			\end{split}
		\ee
	\item On the other hand, by setting $\mu=m,\ \nu=n,\ \rho=p,\ \sigma=0$ in the EoM \eqref{EoMgen} and taking the divergence $\partial_n$, we get
      \begin{align}
	\partial_n\frac{\delta S_{inv}}{\delta A_{mn|p0}}&=a_1\left[-\eta^{mp}\left(\nabla^2F^0+\partial^0\partial_nF^n\right)+\partial^p\left(\partial^mF^0+\partial^0F^m\right)\right]+\label{DEoM-mn}\\
	&\quad+a_2\left\{\eta^{mp}\partial_\alpha\partial_nF^{\alpha n0}-\partial_\alpha\partial^p F^{\alpha m0}+\tfrac1 3 \left[\partial^m\partial_n\left(F^{0np}-F^{pn0}\right)+\right.\right.\nonumber\\
	&\left.\left.\quad+\nabla^2\left(F^{pm0}-F^{0mp}\right)+\partial^p\partial_n\left(F^{nm0}-F^{mn0}\right)+\partial^0\partial_n\left(F^{mnp}-F^{nmp}\right)\right]\right\}+\nonumber\\
	&\quad+4a_3\partial_\alpha F^{\alpha mn|p0}\ ,\nonumber
	\end{align}
	where $\nabla^2\equiv\partial_i\partial^i$. By symmetrising the remaining free indices $m$ and $p$ of \eqref{DEoM-mn} we find
	    \be
	    \partial_n\frac{\delta S_{inv}}{\delta A_{mn|p0}}+\partial_n\frac{\delta S_{inv}}{\delta A_{pn|m0}}=\partial^0\partial_a\left(\Pi^{am|p0}+\Pi^{an|p0}\right)+\partial_a\partial_b\Pi^{ma|pb}\ ,
	    \ee
	    where we noticed that $\partial_a\partial_b(F^{amb|n0}+F^{anb|m0})=-\partial_a\partial_bF^{0ma|nb}$. Going on-shell the first term vanishes due to the EoM \eqref{EoMm0n0}, leaving thus
	    \be
	    \partial_a\partial_b\Pi^{ma|nb}=0\ ,\label{gauss-gen}
	    \ee
	which is a Gauss-like constraint for the conjugate momentum of the whole theory. 
	\ei
We can therefore say that a fractonic behaviour is already present at the level of the general theory, independently from the value of the constants $a_1,\ a_2,\ a_3$, in terms of the generalised Gauss law \eqref{gauss-gen} emerging from the EoM of the action $S_{inv}$ \eqref{Sinv}. This Gauss law is of the same kind as the one appearing in \cite{Pai:2018qnm}, introduced to describe fractonic-line excitations. Moreover, we observed in \eqref{traceless1} and \eqref{traceless2} that there exist some combinations of the parameters $a_1,\ a_2,\ a_3$ implying tracelessness of the conjugate momentum and of the theory itself (\emph{i.e.} the theory does not depend on either $A_{\mu\nu}(x)$ and/or $A(x)$). This can be the sign of additional fractonic features; in fact in the rank-2 models \cite{Pretko:2016kxt,Pretko:2016lgv} tracelessness conditions on the electric tensor field, \emph{i.e.} conjugate momentum, imply additional constraints, such as quadrupole conservation, which further limit the motion of the quasiparticles, thus giving rise to other kinds of excitations such as lineons and planons \cite{Pretko:2016kxt,Pretko:2016lgv,Pretko:2020cko,Seiberg:2020wsg,Gorantla:2022pii,Figueroa-OFarrill:2025yum}. Therefore the existence of combinations of parameters allowing for a traceless $\Pi^{\mu\nu|\rho\sigma}(x)$ \eqref{Pi-tot}, in addition to the general Gauss law \eqref{gauss-gen}, could hint towards a family of new traceless rank-4 fracton behaviours and new fractonic ``extended'' excitations, which will be worth investigating in the future. However, at the level of the full theory, \emph{i.e.} for generic coefficients $a_{1,2,3}$, the electromagnetic analogy ends here, as no Amp\`ere- or Faraday-like laws can be easily identified.

\subsection{Area-metric gravity and linearised metric-induced ansatz}\label{section-area-metric}
Area-metric geometry \cite{Schuller:2005yt, Punzi:2006nx,Borissova:2023yxs} is a natural generalisation of Riemannian geometry in which the fundamental spacetime structure is not a rank-2 metric $g_{\mu\nu}(x)$, but a rank-4 tensor $G_{\mu\nu|\rho\sigma}(x)$ satisfying
	\begin{equation}
	G_{\mu\nu|\rho\sigma}
	=
	G_{\rho\sigma|\mu\nu}
	\qquad;\qquad
	G_{\mu\nu|\rho\sigma}
	=
	- G_{\nu\mu|\rho\sigma}
	=
	- G_{\mu\nu|\sigma\rho}\ ,
	\end{equation}
together with the cyclicity (first Bianchi-type) identity
	\begin{equation}
	G_{\mu\nu|\rho\sigma}
	+
	G_{\mu\rho|\sigma\nu}
	+
	G_{\mu\sigma|\nu\rho}
	=
	0\ ,
	\end{equation}
and a nondegeneracy condition
	\begin{equation}
	\det(G_{\mu\nu|\rho\sigma}) \ne 0\ ,
	\end{equation}
where $G_{\mu\nu|\rho\sigma}(x)$ is thought of as a $\binom d2\times\binom d2$ matrix in $d$ spacetime dimensions. These symmetries and conditions ensure that $G_{\mu\nu|\rho\sigma}(x)$ defines a symmetric bilinear form on the space of two-forms. In contrast to the standard metric, which assigns lengths to vectors, an area-metric assigns norms directly to antisymmetric surface elements. For this reason, it provides a natural geometric framework whenever extended objects or flux degrees of freedom play a fundamental role.
In ordinary Riemannian geometry, an area-metric is not an independent structure but is induced by a metric $g_{\mu\nu}(x)$ via
	\begin{equation}\label{Gg}
	G^{(g)}_{\mu\nu|\rho\sigma}
	=
	g_{\mu\rho}g_{\nu\sigma}
	-
	g_{\mu\sigma}g_{\nu\rho}\ .
	\end{equation}
However, in general one may consider $G_{\mu\nu|\rho\sigma}(x)$ as fundamental and not necessarily reducible to a metric-induced form. Such non-metric area geometries appear in effective descriptions of electrodynamics in media \cite{Hehl,Schuller:2009hn}, pre-metric gravity formulations \cite{Ho:2015cza}, and string-inspired extensions of spacetime structure \cite{Borissova:2024cpx}, where the propagation of gauge fields is governed by an effective constitutive tensor of area-metric type.
A dynamical theory of area-metric gravity promotes $G_{\mu\nu|\rho\sigma}(x)$ to a spacetime field. Linearising around Minkowski space
	\begin{equation}\label{GA}
	G_{\mu\nu|\rho\sigma}
	=
	\eta_{\mu\rho}\eta_{\nu\sigma}
	-
	\eta_{\mu\sigma}\eta_{\nu\rho}
	+
	A_{\mu\nu|\rho\sigma}\ ,    
	\end{equation}
the fluctuation $A_{\mu\nu|\rho\sigma}(x)$ inherits precisely the algebraic symmetries of an area-metric. We therefore observe that the rank-4 gauge field introduced in the previous Sections admits a natural geometric interpretation as the linearised fluctuation of an area-metric background.
This interpretation clarifies the structure of the gauge symmetry \eqref{symm}, which can be viewed as the linearised remnant of longitudinal diffeomorphism-like transformations acting on the underlying area geometry, and, as in its rank-2 counterpart \cite{Bertolini:2022ijb,Pretko:2016lgv} strongly ties fractonic models with gravity-like theories, which therefore seems an underlying recurring feature. Notice that the above gauge transformation \eqref{symm} can be tied to an area-metric diffeomorphism-like transformation and area geometry, but \textit{a priori} it has no relation to the usual infinitesimal diffeomorphisms, unless the induced metric ansatz \eqref{Gg} is taken into account. On the other hand one can think about the usual knowledge of diffeomorphisms as a smooth bijection between manifolds. Infinitesimally a diffeomorphism reduces to a vector field. This is the case for area-metric theories as well, $i.e.$ the notion of a diffeomorphism has not changed. What has changed is the relationship between diffeomorphisms and gauge symmetries. In Einstein gravity, the gauge symmetries are the same as diffeomorphisms, which are integrated versions of vector fields. In metric-induced area-metric gravity, gauge transformations no longer agree with diffeomorphisms (see $e.g.$ \cite{Borissova:2023yxs}  Eq. (3.8)); there are additional gauge transformations that do not correspond to diffeomorphisms, although there are some gauge transformations that arise from linearised diffeomorphisms. Moreover, for non-linearised area metrics, there exists a suitable notion of Killing vectors defined exactly the same way as ordinary (pseudo-)Riemannian geometry. In (pseudo-)Riemannian geometry, a Killing vector is an infinitesimal isometry, $i.e.$ a vector field $X$ such that $\mathcal L_Xg=0$, where $g(x)$ is the metric tensor (of rank two) and $\mathcal L$ is the Lie derivative of a tensor field with respect to a vector field. Similarly, in the case of area-metrics, a Killing vector is a vector field $X$ such that $\mathcal L_X\omega=0$, where $\omega$ is the area-metric, which is a rank-four tensor field. Now, linearisation changes the Killing vectors visible in the theory. Recall that, in Einstein gravity, if we linearise atop a background metric $g_0(x)$ as $g(x)=g_0(x)+h(x)$, then the Killing vectors of $g_0(x)$ survive as global spacetime symmetries of the linearised theory; in particular, if we linearise atop Minkowski metric, then we see the Killing vectors of Minkowski space ($i.e.$ Poincaré algebra) as symmetries of linearised Einstein gravity. Similarly, we can linearise the non-fractonic area-metric gravity on a background area metric $\omega_0(x)$ (with or without fractonic terms); then the Killing vectors of $\omega_0(x)$ survive as global spacetime symmetries of the linearised theory. In this paper, we only consider the (area-metric analogue of) Minkowski metric, so this simply recovers the Poincaré algebra.\\

A particularly important subsector is obtained when the area-metric fluctuation is \emph{metric-induced}. Expanding a metric
	\begin{equation}
	g_{\mu\nu}=\eta_{\mu\nu}+h_{\mu\nu}\ ,
	\end{equation}
and inserting it into the metric-induced area metric $G^{(g)}_{\mu\nu|\rho\sigma}(x)$ \eqref{Gg} and comparing with \eqref{GA}, one finds at linear order
	\begin{equation}\label{eq:A-metric-induced}
	\tilde A_{\mu\nu|\rho\sigma}
	=
	\eta_{\nu\sigma}h_{\mu\rho}
	+
	\eta_{\mu\rho}h_{\nu\sigma}
	-
	\eta_{\nu\rho}h_{\mu\sigma}
	-
	\eta_{\mu\sigma}h_{\nu\rho}\ .
	\end{equation}
Thus the symmetric rank-2 field $h_{\mu\nu}(x)$ generates a constrained subsector of the full rank-4 theory. As we shall see, this metric-induced subsector reproduces linearised gravity together with a covariant fracton theory \cite{Bertolini:2022ijb} for appropriate choices of the invariant couplings, whereas the non-metric sector gives rise to a higher-rank Maxwell-like theory with fractonic behaviours. In this sense, fracton physics for extended objects emerges naturally from the non-metric degrees of freedom of area-metric gauge theory. To show the connection between our rank-4 model \eqref{Sinv} in the metric-induced case with the rank-2 theory of linearised gravity and covariant fractons described in \cite{Bertolini:2022ijb}, we consider the spin-2 gauge field $h_{\mu\nu}(x)$ to transform only under longitudinal diffeomorphisms, \emph{i.e.}
	\begin{equation}
	\delta^{(\textit{long})} h_{\mu\nu} =\partial_\mu \partial_\nu \lambda\ ,\label{eq:longitudinal-diffeo}
	\end{equation}
with $\lambda(x)$ a scalar gauge parameter, which is the gauge transformation that generates rank-2 fractonic behaviours. The field strength depending on $h_{\mu\nu}(x)$ and invariant under longitudinal diffeomorphisms \eqref{eq:longitudinal-diffeo} is \cite{Bertolini:2022ijb}
	\be\label{F2}
	\mathcal{F}_{\mu\nu\rho}=\mathcal{F}_{\nu\mu\rho}\equiv\partial_\mu h_{\nu\rho}+\partial_\nu h_{\rho\mu}-2\partial_\rho h_{\mu\nu}\ .
	\ee
From  \eqref{eq:A-metric-induced}, using \eqref{eq:longitudinal-diffeo}, we obtain
	\be
	\delta^{(\textit{long})}\tilde A_{\mu\nu|\rho\sigma}= \eta_{\nu\sigma}\partial_\mu\partial_\rho\lambda+ \eta_{\mu\rho}\partial_\nu\partial_\sigma\lambda	-\eta_{\nu\rho} \partial_\mu\partial_\sigma\lambda	- \eta_{\mu\sigma}\partial_\nu\partial_\rho\lambda	\ ,
	\label{eq:deltaA-longitudinal}
	\ee
thus the full (original) gauge transformation \eqref{symm} of the rank-4 field matches the longitudinal, linearised approximation \eqref{eq:deltaA-longitudinal} if
	\be
	\lambda_{\mu\nu}=-\eta_{\mu\nu}\lambda\ .
	\ee
Concerning the field strength $F_{\alpha\mu\nu|\rho\sigma}(x)$, using \eqref{eq:A-metric-induced} and applying its definition \eqref{F}, we get 
	\begin{equation}\label{F-explicit}
		\begin{split}
		\tilde F_{\alpha\mu\nu|\rho\sigma}&=\eta_{\mu\rho}\left(2\partial_\alpha h_{\nu\sigma}-\partial_\nu h_{\alpha\sigma}-\partial_\sigma h_{\alpha\nu}\right)-\eta_{\mu\sigma}\left(2\partial_\alpha h_{\nu\rho}-\partial_\nu h_{\alpha\rho}-\partial_\rho h_{\alpha\nu}\right)+\\
		&\quad+\eta_{\nu\sigma}\left(2\partial_\alpha h_{\mu\rho}-\partial_\mu h_{\alpha\rho}-\partial_\rho h_{\alpha\mu}\right)-\eta_{\nu\rho}\left(2\partial_\alpha h_{\mu\sigma}-\partial_\mu h_{\alpha\sigma}-\partial_\sigma h_{\alpha\mu}\right)+\\
		&\quad+\eta_{\alpha\mu}\left(\partial_\sigma h_{\nu\rho}-\partial_\rho h_{\nu\sigma}\right)-\eta_{\alpha\nu}\left(\partial_\sigma h_{\mu\rho}-\partial_\rho h_{\mu\sigma}\right)+\\
		&\quad+\eta_{\alpha\rho}\left(\partial_\nu h_{\mu\sigma}-\partial_\mu h_{\nu\sigma}\right)-\eta_{\alpha\sigma}\left(\partial_\nu h_{\mu\rho}-\partial_\mu h_{\nu\rho}\right)\ ,
		\end{split}
	\end{equation}
which can be rewritten, together with its contractions \eqref{Fmunu}, \eqref{Fmu}, in terms of the $\mathcal{F}$-field strength \eqref{F2} as
	\begin{align}
	\tilde F_{\alpha\mu\nu|\rho\sigma}&=-\eta_{\mu\rho}\mathcal{F}_{\nu\sigma\alpha}+\eta_{\mu\sigma}\mathcal{F}_{\nu\rho\alpha}-\eta_{\nu\sigma}\mathcal{F}_{\mu\rho\alpha}+\eta_{\nu\rho}\mathcal{F}_{\mu\sigma\alpha}-\\
	&\quad-\tfrac{1}{3}\left[\eta_{\alpha\mu}\left(\mathcal{F}_{\nu\rho\sigma}-\mathcal{F}_{\nu\sigma\rho}\right)-\eta_{\alpha\nu}\left(\mathcal{F}_{\mu\rho\sigma}-\mathcal{F}_{\mu\sigma\rho}\right)+\eta_{\alpha\rho}\left(\mathcal{F}_{\sigma\mu\nu}-\mathcal{F}_{\sigma\nu\mu}\right)-\eta_{\alpha\sigma}\left(\mathcal{F}_{\rho\mu\nu}-\mathcal{F}_{\rho\nu\mu}\right)\right]\nonumber\\
	\tilde F_{\alpha\mu\nu}&=-\mathcal{F}_{\mu\nu\alpha}-\eta_{\mu\nu}\mathcal{F}^\lambda_{\ \lambda\alpha}+\tfrac1 2 \eta_{\alpha\mu}\mathcal{F}^\lambda_{\ \lambda\nu}+\tfrac1 2 \eta_{\alpha\nu}\mathcal{F}^\lambda_{\ \lambda\mu}\\
	\tilde F_\alpha&=-4\mathcal{F}^\lambda_{\ \lambda\alpha}\ .
	\end{align}
The invariant action $S_{inv}$ \eqref{Sinv}, when linearised according to \eqref{eq:A-metric-induced}, gives
	\be
		\begin{split}
		\tilde S_{inv}[h]&=\int d^4x\left(-\tfrac{a_1}{4}\tilde F_\alpha \tilde F^\alpha	-\tfrac{a_2}{6}\tilde F_{\alpha\mu\nu}\tilde F^{\alpha\mu\nu}-\tfrac{a_3}{8}\tilde F_{\alpha\mu\nu|\rho\sigma}\tilde F^{\alpha\mu\nu|\rho\sigma}\right)\\
		&=\int d^4x \left[-\tfrac2 3\left(a_3+\tfrac{a_2}{4}\right)\mathcal{F}_{\mu\nu\rho}\mathcal{F}^{\mu\nu\rho}-\left(4a_1+\tfrac5 4 a_2+a_3\right)\mathcal{F}^\lambda_{\ \lambda\alpha}\mathcal{F}_\gamma^{\ \gamma\alpha}\right]\\
		&=g_{\textsc{lg}}S_{\textsc{lg}}+g_\textit{fr}S_\textit{fr}\ ,
		\end{split}
	\ee
with
	\be
	g_{\textsc{lg}}\equiv16a_1+5a_2+4a_3\qquad;\qquad g_\textit{fr}\equiv2(8a_1+3a_2+4a_3)\ ,
	\ee
which coincide with the general invariant action of \cite{Bertolini:2022ijb} describing a mixing of linearised gravity and fractonic contributions. In particular one can isolate linearised gravity or fractons by setting $g_\textit{fr}=0$ and $g_{\textsc{lg}}=0$ respectively.

\section{A Maxwell-like theory for string-like fractons}\label{sec3}
In analogy with \cite{Bertolini:2022ijb}, in order to isolate the Maxwell-like term and identify the covariant fracton theory, we now set $a_1=a_2=0$ and $a_3\equiv a$ in the invariant action $S_{inv}$ \eqref{Sinv}, and study the higher-rank Maxwell-like theory given by
	\be\label{S}
	S_\textit{fr}=-\tfrac{a}{8}\int d^4x F_{\alpha\mu\nu|\rho\sigma}F^{\alpha\mu\nu|\rho\sigma}\ .
	\ee
From \eqref{Pi-tot}, setting $a_1=a_2=0$, we see that the conjugate momentum of the above action is
	\be\label{pi}
	\Pi^{\mu\nu|\rho\sigma}=\frac{\delta S_\textit{fr}}{\delta\partial_t A_{\mu\nu|\rho\sigma}}=-4aF^{0\mu\nu|\rho\sigma}\ ,
	\ee
and similarly, from \eqref{EoMgen}, we obtain the following Maxwell-like EoM
	\be
	\frac{\delta S_\textit{fr}}{\delta A_{\mu\nu|\rho\sigma}}=4a\partial_\alpha F^{\alpha\mu\nu|\rho\sigma}\ .\label{eom}
	\ee
From the general case of Section \ref{section-general eom} we can already say that a Gauss-like constraint exists, being \eqref{gauss-gen}, however we will see that the theory described by the action $S_\textit{fr}$ \eqref{S} contains additional electromagnetic-like and fractonic features, and will allow for a full fractonic string-like interpretation.

\subsection{Generalised electromagnetism}\label{section-em-like}
We now analyse the on-shell components of the EoM \eqref{eom}. Among the indices $\mu,\nu,\rho,\sigma$, due to symmetry, at most two can be directed along time. Thus, we have the following three cases, depending on whether there are two, one, or no time components among the indices.
	\bi
	\item $\pmb{\mu=m,\ \nu=0,\ \rho=n,\ \sigma=0}$ : in this case, the nonvanishing components of \eqref{eom} are
		\be\label{eom00}
			\begin{split}
			0&=4a\partial_\alpha F^{\alpha m0|n0}=4a\partial_0 \cancel{F^{0 m0|n0}}+4a\partial_a F^{a m0|n0}\\
			&=4a\partial_a\left(2\partial^aA^{m0|n0}-\partial^mA^{n0|a0}+\partial^0A^{n0|am}-\partial^nA^{m0|a0}+\partial^0A^{m0|an}\right)\ .
			\end{split}
		\ee
A particular solution is given by
		\be\label{sol}
		A^{m0|n0}=\partial^0\phi^{mn}\quad;\quad A^{mn|p0}=-\partial^m\phi^{np}+\partial^n\phi^{mp}\ ,
		\ee
with $\phi^{mn}(x)=\phi^{nm}(x)$ a symmetric rank-2 field. This solution can also be written in a more compact way as
		\be
		A^{m\mu|n0}=\delta^m_p\partial^\mu\phi^{np}-\delta^\mu_p\partial^m\phi^{np}\ .
		\ee
A symmetric tensor field of this form also appears in the noncovariant rank-4 theory of \cite{Pai:2018qnm}, in which it is introduced by hand as a Lagrange multiplier, in order to enforce a Gauss law and therefore impose a fractonic behaviour on the theory. In our case the existence of the rank-2 symmetric tensor $\phi^{mn}(x)$ naturally appears as a solution of the EoM \eqref{eom00}, without any \emph{ad hoc} implementation. As we shall see, this solution will constitute a sort of fractonic embedding of our covariant rank-4 model. Notice that the gauge choice $\lambda_{\mu0}(x)=0$ allows us to easily identify the transformation of the rank-2 field to be
		\be
		\delta\phi^{mn}=-\partial^0\lambda^{mn}\ , 
		\ee
which is the same as that in \cite{Pai:2018qnm}. We will  keep this solution \eqref{sol} in the remaining analysis of the theory. In this case the field strength also have the following properties
		\begin{align}
		F^{amn|p0}&=\tfrac1 2 F^{0mn|ap}\label{Fsol1}\\
		F^{0mn|p0}&=0\ .\label{Fsol2}
		\end{align}
	\item $\pmb{\mu=m,\ \nu=n,\ \rho=p,\ \sigma=0}$ :
		\be
		0=4a\partial_\alpha F^{\alpha mn|p0}=4a\partial_0 \cancel{F^{0 mn|p0}}+4a\partial_a F^{a mn|p0}=2a\partial_aF^{0mn|ap}=-\tfrac{1}{2}\partial_a\Pi^{mn|ap}\ ,\label{eom0i}
		\ee
where we used the properties \eqref{Fsol1} and \eqref{Fsol2} and the definition of conjugate momentum \eqref{pi}. This is a more general higher-rank Gauss constraint than the one found for the full theory \eqref{gauss-gen}, associated to a rank-4 electric field
		\be\label{E}
			\begin{split}
			E^{mn|pq}&\equiv\Pi^{mn|pq}=-4aF^{0mn|pq}\\
			&=8a\left(-\partial^0A^{mn|pq}+\partial^m\partial^p\phi^{nq}-\partial^m\partial^q\phi^{np}+\partial^n\partial^q\phi^{mp}-\partial^n\partial^p\phi^{mq}\right)\ ,
			\end{split}
		\ee
which matches exactly the one defined in \cite[eq.\ (10)]{Pai:2018qnm}. By computing $\partial_m$ of \eqref{eom0i} we also obtain
		\be\label{gauss1}
		\partial_m\partial_pE^{mn|pq}=0\ ,
		\ee
which is the same of \cite[eq.\ (13)]{Pai:2018qnm} in vacuum, and the starting point for deriving fractonic string-like excitations, as we shall see.
	\item $\pmb{\mu=m,\ \nu=n,\ \rho=p,\ \sigma=q}$ : 
		\be\label{ampere1}
			\begin{split}
			0&=4a\partial_\alpha F^{\alpha mn|pq}
			=4a\partial_0 F^{0 mn|pq}+4a\partial_a F^{a mn|pq}\\
			&=- \left[\partial_0E^{mn|pq}+\tfrac{2a}{k}\left(\epsilon^{0amn}\partial_aB^{pq}+\epsilon^{0apq}\partial_aB^{mn}\right)\right]\ ,
			\end{split}
		\ee
where we have defined the antisymmetric magnetic tensor field
		\be\label{B}
		B^{mn}=-B^{nm}\equiv k\epsilon_{0ijk}\partial^iA^{jk|mn}=\tfrac{k}{4}\epsilon_{0abc}F^{abc|mn}
		\ee
 which is the same as the one of  \cite[eq.\ (10)]{Pai:2018qnm} (with $k$ a generic constant later to be tuned), and in complete analogy with the standard electromagnetic theory. Indeed we also have
		\be
		F^{a mn|pq}=-\tfrac{1}{2k}\left(\epsilon^{0amn}B^{pq}+\epsilon^{0apq}B^{mn}\right)\ .
		\ee
From \eqref{ampere1} we thus get an Amp\`ere-like equation
		\be
		\partial_0E^{mn|pq}+\tfrac{2a}{k}\left(\epsilon^{0amn}\partial_aB^{pq}+\epsilon^{0apq}\partial_aB^{mn}\right)=0\ ,\label{ampere}
		\ee
which coincides with \cite[eq.\ (15)]{Pai:2018qnm} in vacuum.
	\ei
We have therefore found from the EoM \eqref{eom} of our action $S_\textit{fr}$ \eqref{S} a first set of Maxwell-like equations, namely the Gauss constraints \eqref{eom0i} and \eqref{gauss1}, and the  Amp\`ere-like law \eqref{ampere}, written in terms of generalised electric and magnetic tensor fields $E^{mn|pq}(x)$ \eqref{E}  and $B^{mn}(x)$  \eqref{B} depending on the field strength $F^{\alpha\mu\nu|\rho\sigma}(x)$ \eqref{F} in an electromagnetic-like fashion, thus invariant at sight. One difference from standard electromagnetism comes from the absence of a vanishing divergence for the magnetic field, in fact
	\be
	\partial_aB^{ab}=k\epsilon_{0ijk}\partial^i\partial_aA^{jk|ab}\neq0\ .
	\ee
	 To follow our electromagnetic analogy, we can now look at the Bianchi-like identity \eqref{bianchi}, in particular at $\alpha=a,\ \beta=b$
	\be
		\begin{split}
		0&=\epsilon^{\mu\nu\rho\sigma}\partial_\mu F_{\nu ab|\rho\sigma}
		=\epsilon^{0mnp}\partial_0F_{mab|np}-2\epsilon^{0mnp}\partial_mF_{0ab|np}\\
		&=-\tfrac{4}{k}\partial_0B_{ab}-\tfrac{1}{2a}\epsilon^{0mnp}\partial_mE_{ab|np}\ ,
		\end{split}
	\ee
where we used the property \eqref{Fsol1} of the solution \eqref{sol}, and the definitions of the electric and magnetic tensor fields \eqref{E} and \eqref{B}. We thus obtained a higher-rank Faraday-like law
	\be
	\partial_0B_{ab}+\tfrac{k}{8a}\epsilon^{0mnp}\partial_mE_{ab|np}=0\ ,\label{faraday}
	\ee
which, again, is compatible with the one identified in \cite[eq.\ (8)]{Pai:2018qnm}.  To summarise, here from the EoM of the action $S_\textit{fr}$ \eqref{S} and Bianchi-like identity \eqref{bianchi} we found the following Maxwell-like equations
	\begin{align}
	\partial_mE^{mn|pq}=0\qquad;\qquad\partial_m\partial_pE^{mn|pq}&=0\\
	\partial_0E^{mn|pq}+\tfrac{2a}{k}\left(\epsilon^{0amn}\partial_aB^{pq}+\epsilon^{0apq}\partial_aB^{mn}\right)&=0\\
	\partial_0B^{ab}-\tfrac{k}{8a}\epsilon_{0mnp}\partial^mE^{ab|np}&=0\ ,
	\end{align}
for a rank-4 electric field $E^{mn|pq}(x)$ \eqref{E} and rank-2 antisymmetric magnetic field $B^{mn}(x)$ \eqref{B}, defined in terms of the invariant field strength $F^{\alpha\mu\nu|\rho\sigma}(x)$ in complete analogy to the standard electromagnetic theory, and compatibly with the fractonic model of \cite{Pai:2018qnm}, but with the relevant difference that here all the results originates from first principles, with no \emph{ad hoc} requirement. We can finally rewrite the action \eqref{S} in terms of the electric and magnetic tensor fields \eqref{E} and \eqref{B}:
	\be\label{SEB}
		\begin{split}
		S_\textit{fr} &=-\tfrac{a}{8}\int d^4x F_{\alpha\mu\nu|\rho\sigma}F^{\alpha\mu\nu|\rho\sigma}
		 =-\tfrac{a}{8}\int d^4x \left(2F_{0mn|pq}F^{0mn|pq}+F_{amn|pq}F^{amn|pq}\right)\\
		 &=\tfrac{1}{64a}\int d^4x \left(E_{mn|pq}E^{mn|pq}-\tfrac{32a^2}{k^2}B^{mn}B_{mn}\right)\ ,
		\end{split}
	\ee
where we used the properties \eqref{Fsol1} and \eqref{Fsol2} of the solution \eqref{sol}. Indeed the above action can be tuned to be of the form $E^2-B^2$, and thus the fractonic electromagnetic analogy holds.

\subsection{Coupling to fractonic brane-like sources}\label{section-sources}
In order to properly identify the fractonic excitation that characterise our model, their behaviour, and in particular the constraints they are subject to through the Gauss-like laws, we couple the action $S_\textit{fr}$ \eqref{S} to a source term, for which  the total action is
	\be\label{Stot}
	S_{tot}=S_\textit{fr}+S_{J}=S_\textit{fr}+\frac1 4 \int d^4x A_{\mu\nu|\rho\sigma}J^{\mu\nu|\rho\sigma}\ ,
	\ee
with $J^{\mu\nu|\rho\sigma}(x)$ a current with the same index structure of $A_{\mu\nu|\rho\sigma}(x)$. This should be seen as a smeared version of worldvolume couplings to branes. Note that, since our gauge field $A_{\mu\nu|\rho\sigma}(x)$ is not a differential form, the rank of the differential form does not automatically determine the dimension of the brane that it couples to. Instead, for instance, consider a $p$-brane that is equipped with a worldvolume area metric $G^{\textsc{ab}|\textsc{cd}}(x)$ (where $\textsc{a,b,...}=\{0,... ,p\}$ are worldvolume indices), which, for instance, might be induced by a worldvolume metric $g^{\textsc{ab}}(x)$ as $G^{\textsc{ab}|\textsc{cd}}=\sqrt{-g}\left(g^{\textsc{ac}}g^{\textsc{bd}}-g^{\textsc{ad}}g^{\textsc{bc}}\right)$. Then one can write the following coupling term in the brane worldvolume action
	\be
	\int d^py\,g^{\textsc{ab}|\textsc{cd}}A_{\mu\nu|\rho\sigma}\partial_\textsc{a} X^\mu\partial_\textsc{b} X^\nu\partial_\textsc{c} X^\rho\partial_\textsc{d} X^\sigma\ ,
	\ee
where $X(y)$ is the embedding map from the brane worldvolume into spacetime. Note that this is an area-metric analogue of the Polyakov action for a brane \cite{Borissova:2024cpx}.
In that case, the current $J^{\mu\nu|\rho\sigma}(x)$ will be
	\be
	J^{\mu\nu|\rho\sigma}= \int d^py\,\delta(x-X(y))g^{\textsc{ab}|\textsc{cd}}\partial_\textsc{a} X^\mu\partial_\textsc{b} X^\nu\partial_\textsc{c} X^\rho\partial_\textsc{d} X^\sigma\ .
	\ee
Notice that, as a consequence of the gauge symmetry \eqref{symm}, from the total action $S_{tot} \eqref{Stot}$ we can identify the following conserved-current equation
	\be\label{ddJ}
	\partial_\mu\partial_\rho J^{\mu\nu|\rho\sigma}=0\ .
	\ee
Given the addition of the source term $S_J$ in \eqref{Stot}, the EoM \eqref{eom} now read
	\be\label{S-brane}
	\frac{\delta S_{tot}}{\delta A_{\mu\nu|\rho\sigma}}=4a\partial_\alpha F^{\alpha\mu\nu|\rho\sigma}+J^{\mu\nu|\rho\sigma}\ .
	\ee
The on-shell components of the above EoM \eqref{S-brane} now are as follows.
	\bi
	\item $\pmb{\mu=m,\ \nu=0,\ \rho=n,\ \sigma=0}$ : for which \eqref{S-brane} reduces to
		\be
		0=4a\partial_a F^{a m0|n0}+J^{m0|n0}\ ,
		\ee
which in the fractonic solution \eqref{sol} that defines the electric tensor field $E^{mn|pq}(x)$ \eqref{E} and identifies the electromagnetic analogy, implies
		\be\label{Jm0n0=0}
		J^{m0|n0}=0\ .
		\ee
	\item $\pmb{\mu=m,\ \nu=n,\ \rho=p,\ \sigma=0}$  : 
		\be
		-\tfrac1 2 \partial_a E^{mn|ap}+J^{mn|p0}=0\qquad\Rightarrow\qquad\partial_a E^{mn|ap}=2d^{mn|p}\ ,\label{gauss-vec-mat}
		\ee
where we have defined the rank-3 charge
		\be\label{d}
		d^{mn|p}=-d^{nm|p}\equiv J^{mn|p0}\ ,
		\ee
and by computing $\partial_m$ we get the Gauss law
		\be\label{gauss-mat}
		\partial_m\partial_pE^{mn|pq}=\rho^{nq}\ ,
		\ee
which matches with the one in \cite{Pai:2018qnm} if we define the symmetric rank-2 charge
		\be\label{rho}
		\rho^{mn}=\rho^{nm}\equiv\partial_a\left(d^{am|n}+d^{an|m}\right)\ .
		\ee
The above relation between the two charges allows us to interpret $d^{mn|p}(x)$ as a sort of antisymmetric dipole-like density \cite{Ebisu:2024eew} for the rank-2 charge $\rho^{mn}(x)$ as
		\be\label{d-dipole}
		\int dV d^{ab|m}=\tfrac1 3 \int dV\left(x^a\rho^{bm}-x^b\rho^{am}\right)\ .
		\ee
Moreover, as a consequence of the Gauss-like constraints \eqref{gauss-mat} and \eqref{gauss-vec-mat}, the higher‑rank charge densities $\rho^{mn}(x)$ and $d^{mn|p}(x)$ satisfy
		\be
		\partial_m\rho^{mn}=0\quad;\quad\partial_p d^{mn|p}=0\ .
		\ee
The first condition, $\partial_m\rho^{mn}=0$, implies that the symmetric rank‑2 line charge cannot be created or destroyed at isolated points in the bulk, and therefore corresponds to extended, line‑like excitations with restricted mobility -- so‑called \textit{closed fractonic lines} -- as discussed in \cite{Pai:2018qnm}. In this context “closed” means that such line excitations cannot terminate freely in the interior of the system, consistently with the flux conservation laws that lock them in place. Likewise, the additional constraint $\partial_p d^{mn|p}=0$ enforces a generalised tensorial dipole conservation for these extended excitations, further restricting their allowed deformations and motion.  In $d^{mn|p}(x)$, each antisymmetric pair $mn$ defines a polarisation along the string, and the $p$ index indicates the flux direction. Since the strings are closed, there are no endpoints, and the flux must vanish around each loop. 
		\item $\pmb{\mu=m,\ \nu=n,\ \rho=p,\ \sigma=q}$ :  
		\be
		\partial_0E^{mn|pq}+\tfrac{2a}{k}\left(\epsilon^{0amn}\partial_aB^{pq}+\epsilon^{0apq}\partial_aB^{mn}\right)=J^{mn|pq}\ ,\label{ampere-mat}
		\ee
which is an Amp\`ere-like equation associated to a rank-4 current $J^{mn|pq}(x)$. By taking $\partial_m\partial_p$ and using the Gauss-like law \eqref{gauss-mat} we also obtain the following continuity equation
		\be\label{cont}
		\partial_0\rho^{mn}+\partial_a\partial_bJ^{ma|nb}=0\ ,
		\ee
confirmed also by the Ward identity \eqref{ddJ} when using \eqref{Jm0n0=0}.
	\ei
It is interesting to notice that by taking the traces of the Gauss-like constraints \eqref{gauss-vec-mat} and \eqref{gauss-mat} we get vector and scalar versions 
	\be\label{rank-2gauss}
	\partial_nE^{mn}=2d^m\quad;\quad \partial_m\partial_n E^{mn}=\rho\ ,
	\ee
with 
	\be
	E^{mn}=E^{nm}\equiv\eta_{ab}E^{am|bn}\quad;\quad d^m\equiv \eta_{ab}d^{ma|b}\quad;\quad \rho\equiv\eta_{ab}\rho^{ab}\ .
	\ee
These are typical of rank-2 fracton models and lie at the heart of the mobility constraints associated to dipole conservation \cite{Pretko:2016lgv}. In the same way,  the trace of the Amp\`ere-like equation \eqref{ampere-mat} is
	\be
	\partial_0E^{mn}+\tfrac{2a}{k}\left(\epsilon^{0amp}\partial_aB^{n}_{\ p}+\epsilon^{0anp}\partial_aB^{m}_{\ p}\right)=J^{mn}\ ,
	\ee
which again reminds of the one in the scalar charge theory of fractons \cite{Pretko:2016lgv}. Finally we obtain the continuity equation
	\be
	\partial_0\rho+\partial_m\partial_nJ^{mn}=0\ ,
	\ee
with $J^{mn}=J^{nm}\equiv \eta_{ab}J^{am|bn}$, again typical of rank-2 fracton models \cite{Pretko:2016lgv} and encoding dipole conservation \eqref{dipole cons}. Thus it seems that our rank-4 covariant theory also embeds some feature of its rank-2 counterpart, in terms of traces. The trace of $\rho^{mn}$, then, corresponds to a brane coupling to the scalar charge theory of fractons, so that from the perspective of the scalar charge theory a brane source appears as a smeared set of charges, much like a line or plane of uniform charge in ordinary electromagnetism.

\subsection{Fractonic behaviour and constraints}\label{section-constraints}
In the previous section we identified the following Gauss-like equations
	\begin{align}
   	 \partial_a E^{mn|ap}=&2d^{mn|p}\label{gauss-d}\\
	\partial_m\partial_pE^{mn|pq}=&\rho^{nq}\ ,\label{gauss-rho}
	\end{align}
whose traces \eqref{rank-2gauss}, we observed, represent the usual fractonic (rank-2) ones \cite{Pretko:2016lgv}. We now want to analyse these equations with the aim of understanding the conservation laws that they generate, and, ultimately, the constraints on the motion of our extended charges $\rho^{mn}(x)$ and $d^{mn|p}(x)$. Upon integration over an infinite volume (or up to boundary contributions) we trivially observe that both charges $\rho^{mn}(x)$ \eqref{rho} and $d^{mn|p}(x)$ \eqref{d} are conserved. However there are other charge neutrality conditions arising from these Gauss constraints. In particular 
	\bi
	\item from \eqref{gauss-rho} we have that the total dipole momentum has to be conserved
		\be
		\int dVx^k\rho^{mn}=\int dVx^k\partial_a\partial_bE^{am|bn}=-\int dV\partial_aE^{am|kn}=0\ ,
		\ee
which can be similarly obtained from the continuity equation \eqref{cont}.
	\item From \eqref{gauss-d} we see that a component of quadrupole-like momentum (remember that $d^{mn|p}(x)$ is dipole-like due to \eqref{d-dipole}) is conserved
		\be
		\int dVx_jd^{mn|j}=\tfrac1 2\int dVx_j\partial_iE^{mn|ij}=-\tfrac1 2\int dV\eta_{ij}E^{mn|ij}=0\ ,
		\ee
and similarly an angular momentum-like vector
		\be
		\int dV\epsilon_{npq}x^pd^{mn|q}=\tfrac1 2\int dV\epsilon_{npq}x^p\partial_iE^{mn|iq}=-\tfrac1 2\int dV\epsilon_{npq}E^{mn|pq}=0\ ,
		\ee
due to the cyclicity condition \eqref{ciclicity}. Finally the following quantity is also conserved
		\be
			\begin{split}
			\int dV\left(x^kx_jd^{mn|j}+\tfrac1 2 x^2d^{mn|k}\right)&=\tfrac1 2\int dV\left(x_jx^k\partial_iE^{mn|ij}+\tfrac1 2x^2\partial_iE^{mn|ik}\right)\\
			&=\int dV\left(\cancel{\eta_{ij}E^{mn|ij}}-x_jE^{mn|kj}+x_jE^{mn|kj}\right)=0\ .
			\end{split}
		\ee
	\ei
We can therefore summarise that the theory described by the action $S_{tot}$ \eqref{Stot} in the fractonic solution \eqref{sol}, is characterised by the following  constraints
	\begin{align}
	\int dV\rho^{mn}=0\quad&;\quad	\int dV x^k\rho^{mn}=0\label{cons1}\\
	\int dV d^{mn|j}=0\quad&;\quad	\int dVx_jd^{mn|j}=0\label{cons2}\\
	\int dV\epsilon_{0nja}x^ad^{mn|j}=0\quad&;\quad\int dV\left(x^kx_jd^{mn|j}+\tfrac1 2 x^2d^{mn|k}\right)=0\ .\label{cons3}
	\end{align}
These conserved quantities are in complete analogy to those identifying the so-called ``scalar charge theory of fractons'' and ``traceless vector charge theory''  \cite{Pretko:2016lgv}, which, respectively, refers to a scalar charge density $\rho(x)$ and a (dipole-like) vector one $\rho^i(x)$ being of fractonic nature (\emph{i.e.} immobile). In our case the constraints are instead associated to the symmetric rank-2 charge $\rho^{mn}(x)$ \eqref{rho}, and the rank-3 dipole-like density $d^{mn|p}(x)$ \eqref{d}. Therefore following the analogy, these conservations \eqref{cons1}-\eqref{cons3} imply a complete constraint on the motion of both extended charges $\rho^{mn}(x)$ and $d^{mn|p}(x)$, \emph{i.e.}
	\be
	\rho^{mn}\ \to\ \textbf{fractonic}\quad;\quad d^{mn|p}\ \to\ \textbf{fractonic}\ .
	\ee
In fact, despite  \eqref{cons1}-\eqref{cons3}  being higher-rank versions of the constraints appearing in the ``scalar charge theory of fractons'' and ``traceless vector charge theory'', they coincide when we look at them as referred to the components of the charges. Therefore, if, according to \cite{Pretko:2016lgv}, all the components are constrained to be immobile, \emph{i.e.} fractonic, so must be the full tensorial charges. In this regard, a totally immobile fractonic string or brane may be thought of as an infinite number of fractonic particles arranged in a line or surface, similar to how a tensionless string can be regarded as an infinite number of massless particles \cite{Lindstrom:2022iec,Borsten:2025dcx}.

\subsection{Energy-momentum tensor and Lorentz-like force}\label{section-lorentz}
Given that we now have a well defined theory for extended fractons described by the action $S_\textit{fr}$ \eqref{S}, we can now apply the standard covariant calculations in order to compute its energy-momentum tensor, which is
	\be
		\begin{split}
		T_{\alpha\beta}&\equiv\left.-\frac{2}{\sqrt{-g}}\frac{\delta S_\textit{fr}}{\delta g^{\alpha\beta}}\right|_{g^{\alpha\beta}=\eta^{\alpha\beta}}\\
		&=\tfrac{a}{8}\left[-\eta_{\alpha\beta}F_{\lambda\mu\nu|\rho\sigma}F^{\lambda\mu\nu|\rho\sigma}+2\left(F_{\alpha\mu\nu|\rho\sigma}F_{\beta}^{\ \mu\nu|\rho\sigma}+4F^{\mu}_{\ \alpha\nu|\rho\sigma}F_{\mu\beta}^{\quad\nu|\rho\sigma}\right)\right]\ .
		\end{split}
	\ee
The 00-component (\emph{i.e.} energy density) in the fractonic solution \eqref{sol} gives
	\be\label{T00}
		\begin{split}
		T_{00}\equiv u&=\tfrac{a}{8}\left[F_{\lambda\mu\nu|\rho\sigma}F^{\lambda\mu\nu|\rho\sigma}+2\left(F_{0\mu\nu|\rho\sigma}F_{0}^{\ \mu\nu|\rho\sigma}+4F^{\mu}_{\ \,0\nu|\rho\sigma}F_{\mu0}^{\ \ \nu|\rho\sigma}\right)\right]\\
		&=\tfrac{a}{8}\left[F_{\lambda\mu\nu|\rho\sigma}F^{\lambda\mu\nu|\rho\sigma}-2\left(F_{0mn|pq}F^{0mn|pq}+4F_{m0n|pq}F^{m0n|pq}\right)\right]\\
		&=\tfrac{a}{8}\left(F_{\lambda\mu\nu|\rho\sigma}F^{\lambda\mu\nu|\rho\sigma}-4F_{0mn|pq}F^{0mn|pq}\right)\\
		&=-\tfrac{1}{64a}\left(E_{mn|pq}E^{mn|pq}-\tfrac{32a^2}{k^2}B^{mn}B_{mn}\right)+\tfrac{1}{32a}E_{mn|pq}E^{mn|pq}\\
		&=\tfrac{1}{64a}\left(E_{mn|pq}E^{mn|pq}+\tfrac{32a^2}{k^2}B^{mn}B_{mn}\right)\ ,
		\end{split}
	\ee
which, requiring that $a>0$, provides a positive-definite energy density for the theory. Thus, by setting
	\be
	k^2=32a^2\quad\Rightarrow\quad a=\tfrac{1}{4\sqrt2}\abs{k}\ ,\label{k=a}
	\ee
the energy density \eqref{T00} becomes  Maxwell-like, \emph{i.e.}
	\be\label{H}
	T_{00}=\tfrac{1}{64a}\left(E_{mn|pq}E^{mn|pq}+B^{mn}B_{mn}\right)\ ,
	\ee
which, again, reflects the electromagnetic analogy typical of fracton theories \cite{Pretko:2016lgv,Bertolini:2022ijb}. In the same way we can identify the other Maxwell-like components. For the Poynting-like vector we get
	\be
		\begin{split}
		T_{0i}\equiv P_i&=\tfrac{a}{4}\left(F_{0mn|pq}F_i^{\ mn|pq}+4F^m_{\ \ 0n|pq}F_{mi}^{\ \ n|pq}\right)\\
		&=\tfrac{a}{4}F^{0mn|pq}\left(-F_{imn|pq}+2F_{min|pq}\right)\\
		&=-\tfrac{a}{2}F^{0mn|pq}F_{imn|pq}\\
		&=\tfrac{1}{8k}\epsilon_{0imn}E^{mn|pq}B_{pq}\ ,
		\end{split}
	\ee
where we used the property \eqref{cycF2}. Finally the stress tensor is
	\begin{align}
	T_{ij}\equiv\sigma_{ij}&=\tfrac{a}{8}\left[-\eta_{ij}F_{\lambda\mu\nu|\rho\sigma}F^{\lambda\mu\nu|\rho\sigma}+2\left(F_{i\mu\nu|\rho\sigma}F_j^{\ \mu\nu|\rho\sigma}+4F^\mu_{\ i\nu|\rho\sigma}F_{\mu j}^{\ \ \nu|\rho\sigma}\right)\right]\nonumber\\
	&=\tfrac{1}{64a}\eta_{ij}\left(E_{mn|pq}E^{mn|pq}-\tfrac{32a^2}{k^2}B_{mn}B^{mn}\right)+\nonumber\\
	&\quad+\tfrac{a}{4}\left(-8F_{0im|pq}F_{0j}^{\ \ m|pq}+F_{imn|pq}F_j^{\ mn|pq}+4F^m_{\ \ in|pq}F_{mj}^{\ \ n|pq}\right)\nonumber\\
	&=\tfrac{1}{64a}\eta_{ij}\left(E_{mn|pq}E^{mn|pq}-\tfrac{32a^2}{k^2}B_{mn}B^{mn}\right)+\tfrac{a}{4}\left\{-\tfrac{1}{2a^2}E_{im|pq}E_j^{\ m|pq}-\right.\\
	&\quad\left.-\tfrac{1}{4k^2}\eta_{aj}\left[\left(\epsilon_{0imn}B_{pq}+\epsilon_{0ipq}B_{mn}\right)\left(\epsilon^{0amn}B^{pq}+\epsilon^{0apq}B^{mn}\right)+\right.\right.\nonumber\\
	&\quad\left.\left.+4\left(\epsilon_{0min}B_{pq}+\epsilon_{0mpq}B_{in}\right)\left(\epsilon^{0man}B^{pq}+\epsilon^{0mpq}B^{an}\right)\right]\right\}\nonumber\\
	&=\tfrac{1}{64a}\eta_{ij}E_{mn|pq}E^{mn|pq}-\eta_{ij}\tfrac{a}{2k^2}B_{mn}B^{mn}-\tfrac{1}{8a}E_{im|pq}E_j^{\ m|pq}+\tfrac{a}{k^2}\left(\eta_{ij}B_{mn}B^{mn}+2B_{im}B_j^{\ m}\right)\nonumber\\
	&=\tfrac{1}{8a}\left[\tfrac1 8 \eta_{ij}\left(E_{mn|pq}E^{mn|pq}+B_{mn}B^{mn}\right)-E_{im|pq}E_j^{\ m|pq}+\tfrac{1}{2}B_{im}B_j^{\ m}\right]\ ,\nonumber
	\end{align}
which again is Maxwell-like. Notice that in a general $d$-dimensional spacetime we have
	\be
	T^\mu_{\ \mu}=\eta^{\alpha\beta}T_{\alpha\beta}=-\tfrac{a}{8}(d-10)F_{\lambda\mu\nu|\rho\sigma}F^{\lambda\mu\nu|\rho\sigma}\ ,
	\ee
\emph{i.e.} the energy-momentum tensor is traceless only in the case of $d=10$. In the rank-2 covariant case \cite{Bertolini:2022ijb} a similar behaviour existed for $d=6$. This is curious in light of the fact that \(d=6\) and \(d=10\) are the largest spacetime dimensions admitting super-Poincaré algebras with 8 and 16 supercharges respectively and therefore appear in, for instance, the magic pyramid construction \cite{Anastasiou:2013hba} of supergravities.
The tracelessness may appear to be in tension with the usual statement that nonfree conformal field theories, characterised by traceless energy-momentum tensors, can only exist for $d\le6$. However, without matter, the theory we consider is free, and we do not claim that coupling to matter can be conformal. We can now take a look at the divergence of the energy-momentum tensor. For the 0th-component we have
	\begin{align}
	\partial_\mu T^{\mu0}&=\partial_0T_{00}+\partial_iT^{i0}=\partial_0u-\partial_iP^i\nonumber\\
	&=\tfrac{1}{32a}\left(E_{mn|pq}\partial_0E^{mn|pq}+B_{mn}\partial_0B^{mn}\right)-\tfrac{1}{8k}\epsilon_{0imn}\left(\partial^iE^{mn|pq}B_{pq}+E^{mn|pq}\partial^iB_{pq}\right)\label{DT0}\\
	&=\tfrac{1}{32a}\left\{E_{mn|pq}\left[\partial_0E^{mn|pq}+\tfrac{2a}{k}\left(\epsilon^{0imn}\partial_iB^{pq}+\epsilon^{0ipq}\partial_iB^{mn}\right)\right]+B_{mn}\left(\partial_0B^{mn}-\tfrac{k}{8a}\epsilon_{0ipq}\partial^iE^{mn|pq}\right)\right\}\nonumber\\
	&=0\ ,\nonumber
	\end{align}
where we used the Amp\`ere equation \eqref{ampere} in vacuum, and the Faraday-like law \eqref{faraday}. That implies that the total energy of the system is conserved in time, \emph{i.e.}
	\be
	\partial_0 U=\partial_0\int dVu=\int dV\partial_iP^i=0\ ,
	\ee
up to boundary contributions. On the other hand the $i$-component gives
	\begin{align}
	\partial^\mu T_{\mu i}&=-\partial_0T_{0i}+\partial^jT_{ij}=-\partial_0P_i+\partial^j\sigma_{ij}\nonumber\\
	&=-\tfrac{1}{8k}\epsilon_{0imn}\left(B_{pq}\partial_0E^{mn|pq}+E^{mn|pq}\partial_0B_{pq}\right)+\tfrac{1}{8a}\left(\tfrac1 4 E_{mn|pq}\partial_iE^{mn|pq}+\tfrac1 4 B_{mn}\partial_iB^{mn}-\right.\nonumber\\
	&\quad\left.-E^{jm|pq}\partial_jE_{im|pq}-E_{im|pq}\partial_jE^{jm|pq}+\tfrac1 2 B^{jm}\partial_jB_{im}+\tfrac1 2 B_{im}\partial_jB^{jm}\right)\label{DTi}\\
	&=-\tfrac{1}{8k}\left[-\tfrac{2a}{k}\epsilon_{0imn}B_{pq}\left(\epsilon^{0amn}\partial_aB^{pq}+\epsilon^{0apq}\partial_aB^{mn}\right)-\tfrac{k}{8a}\epsilon_{0imn}\epsilon^{0abc}E^{mn|pq}\partial_aE_{pq|bc}\right]+\nonumber\\
	&\quad+\tfrac{1}{8a}\left(\tfrac1 4 E_{mn|pq}\partial_iE^{mn|pq}+\tfrac1 4 B_{mn}\partial_iB^{mn}-E^{jm|pq}\partial_jE_{im|pq}+\tfrac1 2 B^{jm}\partial_jB_{im}+\tfrac1 2 B_{im}\partial_jB^{jm}\right)\nonumber\\
	&=\tfrac{1}{16a}\left(-E^{jm|pq}\partial_jE_{im|pq}+B^{jm}\partial_jB_{im}+\tfrac3 2 B_{im}\partial_jB^{jm}\right)\nonumber\\
	&\equiv T_i^{(\textit{break})}\nonumber\ ,
	\end{align}
where at the third equality we used the Amp\`ere- and Gauss-like laws in vacuum \eqref{ampere} and \eqref{eom0i}, and the Faraday-like law \eqref{faraday}. Therefore the spatial component is not conserved, but broken by $T_i^{(\textit{break})}(x)$. This is a recurrent feature in fracton models, and appears also in the rank-2 cases \cite{Bertolini:2022ijb,Bertolini:2025goo} as a consequence of the breaking of the full diffeomorphisms symmetry (we remind that fractons are invariant only under the longitudinal subgroup  \eqref{eq:longitudinal-diffeo}). By following the same steps as in \eqref{DT0} and \eqref{DTi}, but now considering the fractonic matter coupling $S_J$ in \eqref{Stot}, for which the Gauss and Amp\`ere equations changes as in \eqref{gauss-vec-mat} and \eqref{ampere-mat}, we obtain
	\begin{align} 
	\partial_\mu T^{\mu 0}&=-f^0\\
	\partial_\mu T^{\mu i}&=T^{(\textit{break})i}-f^i\ ,
	\end{align}
which allows us to identify the power density $f^0(x)$ and Lorentz-like force density $f^i(x)$ as
	\begin{align} 
	f^0&\equiv\tfrac{1}{32a}E_{mn|pq}J^{mn|pq}\\
	f^i&\equiv\tfrac{1}{4a}d_{pq|m}E^{im|pq}+\tfrac{1}{16k}J_{mn|pq}\left(\epsilon^{0imn}B^{pq}+\epsilon^{0ipq}B^{mn}\right)\ .\label{lorentz}
	\end{align}
Interpreting $J^{mn|pq}(x)$ as the current associated to the dipole-like extended excitations $d^{mn|p}(x)$ \eqref{d} as
	\be
	J^{mn|pq}\equiv d^{mn|p}v^q-d^{mn|q}v^p+d^{pq|m}v^n-d^{pq|n}v^m\ ,
	\ee
the Lorentz-like force $f^i(x)$ \eqref{lorentz} gets the even more familiar look
	\be
	f^i\equiv\tfrac{1}{4}d_{pq|m}\left[\tfrac{1}{a}E^{im|pq}+\tfrac{1}{k}v_n\left(\epsilon^{0imn}B^{pq}+\epsilon^{0ipq}B^{mn}\right)\right]\ ,
	\ee
as the force acting on the dipole-like extended charge $d^{mn|p}(x)$ \eqref{d}.

\section{Conclusions and Outlook}\label{sec-conclusion}
	In this work we built a covariant higher-rank gauge theory that provides a relativistic completion of the fractonic framework for extended objects (strings) introduced in \cite{Pai:2018qnm}, and recovered some of its assumptions form first principles. By extending the noncovariant gauge symmetry of \cite{Pai:2018qnm} to its full covariant counterpart \eqref{symm}, our formulation justifies that construction within  a pure quantum field theoretical perspective and extends it to a fully Lorentz-covariant setting, where fractonic behaviours are embedded into a unified gauge-theory structure. The full theory described by the general invariant action $S_{inv}$ \eqref{Sinv}, and entirely expressed in terms of the rank-5 field strength $F^{\alpha\mu\nu|\rho\sigma}(x)$, displays indeed a generalised Gauss law that is the footprint of fractonic behaviours, but the central result of our analysis is the identification of a gauge-invariant Maxwell-like theory, obtained by ``switching off'' the constants $a_1,\ a_2$ in the general action, and providing the full electromagnetic-like fractonic content of the theory. In fact the corresponding action allows for a consistent identification of generalised Maxwell-like equations (Gauss, Amp\`ere and Faraday), written in terms of generalised electric and magnetic tensors $E^{mn|pq}(x)$, $B^{mn}(x)$, together with conserved currents, a Maxwell-like energy-momentum tensor, and Lorentz-like force. Remarkably, our theory not only recovers the equations of \cite{Pai:2018qnm} (which gives the dipole-like conservation for strings \eqref{cons1}), justifies its \textit{ad hoc} implementations (such as the introduction of the Lagrange multiplier $\phi_{mn}(x)$, which in our theory is a solution of an equation of motion), and gives the full electromagnetic-like picture, but it also exhibits an additional Gauss constraint, \eqref{gauss-d}, associated to a new fractonic extended charge $d^{mn|p}(x)$ being the dipole of the string \cite{Ebisu:2024eew}. This new, dipole-like extended charge is subject to a new, fully constraining, set of conservation equations \eqref{cons2}, \eqref{cons3}. Therefore, considering the full set of conserved quantities \eqref{cons1}-\eqref{cons3}, the quasiparticle content of the theory described by the Maxwell-like action $S_\textit{fr}$ \eqref{S} is characterised by
	\bi
	\item an immobile, $i.e.$ fractonic, closed string ;
	\item an immobile , $i.e.$ fractonic, dipole of the string. 
	\ei
Thus, the theory provides a natural covariant origin of mobility restrictions for extended objects, which are not imposed. The computation of the energy-momentum tensor confirms consistency with local conservation and reveals a well-defined Lorentz force $f^i(x)$ acting on the extended dipole-like charge $d^{mn|p}(x)$. This force displays the expected higher-moment dependence. We remark that here we focussed on the discussion of the free theory -- which, being free, has no issues with unitarity given appropriate initial conditions -- on couplings to a background source, similar to how Maxwell theory (a free theory) can be coupled to background currents, and to the subsector that fully describes fractonic strings. This subsector, discussed in Section \ref{sec3}, is reconduced to the model in \cite{Pai:2018qnm}, and is characterised by a positive-definite Hamiltonian \eqref{H}. Moreover the theory represents a natural rank-4 extension of the rank-2 covariant model of \cite{Blasi:2022mbl,Bertolini:2022ijb,Bertolini:2023juh}, thus reinforcing the interpretation of a relativistic embedding of fractonic systems. The standard rank-2 fracton gauge theory is tightly connected to linearised gravity \cite{Pretko:2017fbf,Blasi:2022mbl,Bertolini:2023juh,Afxonidis:2023pdq,Pena-Benitez:2023aat} through its gauge symmetry and tensorial nature. In this paper we prove that that is an intrinsic feature of all kinds of fracton gauge theories, even more exotic ones. Indeed we have shown that the geometric structure of our model naturally connects to linearised area-metric gravity and reduces to linearised rank-2 fracton theory \cite{Blasi:2022mbl,Bertolini:2022ijb,Bertolini:2023juh} once we decompose the rank-4 tensor field in terms of the usual linearised metric tensor $h_{\mu\nu}(x)$. Thus we uncover a geometric interpretation of our model: the rank-4 gauge field admits a natural relation to linearised area-metric gravity, which naturally generalises standard gravity with the area, that plays a more fundamental role than length. It is interesting to compare our construction with the theory analysed in \cite{Chatzistavrakidis:2024dkw}. Similarly to our paper, in \cite{Chatzistavrakidis:2024dkw} a rank-4 tensor gauge field  is considered, which however transforms under a different gauge transformation, depending on a rank-3 gauge parameter (see Eq.~(6.3)), while ours is rank-2 \eqref{symm}. That is key in distinguishing the two models, which indeed differ both in the general action and in the number of dimensions they are built in: 5D for \cite{Chatzistavrakidis:2024dkw} and 4D here. In fact the 4D theory of \cite{Chatzistavrakidis:2024dkw} seems to have no physical degrees of freedom, while in our case they are of fractonic nature. That discrepancy can be viewed in terms of the gauge redundancy given by the different kind of gauge parameters: the rank-3 of \cite{Chatzistavrakidis:2024dkw} implies a higher-number of unphysical degrees of freedom than our rank-2. Moreover in our case the resulting Gauss constraints do not merely eliminate unphysical polarisations, but instead enforce higher-moment charge conservation and fractonic mobility restrictions. Consequently, although both theories involve mixed rank-4 tensors, the physical degrees of freedom and their dynamical interpretation are qualitatively different. Given the results obtained in this paper, several directions deserve further investigation. First, the connection with the four-dimensional elasticity theory implied in \cite{Pai:2018qnm} should be explored in detail. The tensorial structure of the gauge field suggests a direct correspondence with generalised strain tensors, potentially leading to a fully covariant fracton-elasticity duality for extended objects. From a quantum field theoretical perspective a systematic analysis of the gauge-fixing, propagators, degrees of freedom, and pole structures is necessary in order to have the full picture for a well defined theory, and to fully characterise its spectrum. That could also help in identifying the parameter space in which the theory, and all its subsectors, live, similarly to \cite{Blasi:2022mbl,Bertolini:2022ijb} for which certain values of the parameters, corresponding to poles in the propagators, indicate transitions between different theories. For the rank-2 case that distinguished linearised gravity, fractons, and traceless fractons. Analogously, in our model this kind of analysis could point us towards a proper action for linearised area-metric and, possibly, other kinds of fractonic behaviours. In this respect, a complete formulation in terms of area-metric geometry beyond the linearised regime may clarify the geometric origin of fractonic constraints and provide a unified description of generalised gravitational theories and subdimensional dynamics in curved space. To our knowledge there is no notion of an analog of the full diffeomorphisms for the non-linearised area-metric theory yet. In the literature one typically reduces to the induced metric ansatz \eqref{Gg}, and thus to the usual infinitesimal diffeomorphisms. In this paper our main focus was the emerging of fractonic strings from a rank-4 bi-form gauge theory, which leaves the full characterisation of the area-metric sector to a future analysis. The study of propagators mentioned above for instance will help identify the pure area-metric sector and isolate its defining (linearised area-metric diffeomorphisms) symmetry. In analogy to what happens in the rank-2 case \cite{Blasi:2022mbl,Bertolini:2023juh} we expect that the gauge transformation \eqref{symm} would emerge from that ``linearised area-metric diffeomorphisms symmetry'' as its longitudinal projection, while the parent symmetry would be more similar to the one of \cite{Chatzistavrakidis:2024dkw} Eq.(6.3). From there the full non-linearised diffeomorphisms could be inferred through a bottom-up approach, or by means of a full geometrical construction. Moreover interacting theories with higher derivatives generically suffer from issues with unitarity. This will require detailed analysis (as well as specific choices of interaction terms), and the definition of a proper covariant derivative. The interpretation of the extended excitations as genuine string- or brane-like objects also deserves a dedicated treatment. It would be important to determine whether a consistent world-volume action can be derived that reproduces the bulk Lorentz force obtained here. Given the results of the covariant and noncovariant rank-2 cases \cite{Bertolini:2023sqa,Prem:2017kxc,Pretko:2017xar,You:2019bvu,Parasar:2025ngk}, the structure of edge modes remains largely unexplored. The higher-rank nature of the gauge symmetry, could lead to new, nontrivial boundary sectors, possibly with topological characteristics, as a natural generalisation of the point-like fractonic case \cite{Bertolini:2023sqa} and more standard quantum field theoretical setups \cite{Stone:1990iw,Wen:1992vi,Cho:2010rk,Blasi:2010gw,Blasi:2011pf,Amoretti:2012hs,Cirio:2013dxa,Amoretti:2014kba,Amoretti:2014iza,Chen:2015gma,Palumbo:2016nku,Geiller:2019bti,Bertolini:2020hgr,Bertolini:2021iku,Bertolini:2022sao,Bertolini:2023wie}. Finally, the inclusion of generalised theta-terms or topological couplings could enrich the phase structure of the model, potentially leading to new types of fractonic response phenomena.	Overall, the present construction provides a covariant framework for fractonic gauge theories, bridging higher-rank tensor gauge fields, area-metric structures, and extended charged objects within a single consistent field-theoretic description.  Our results open a route toward a fully quantum field theory-based description of fractonic strings, connecting higher-rank gauge theories and generalised gravitational structures.

\section*{Acknowledgements}
We thank Nicola Maggiore and Arash Ranjbar for enlightening discussions. This work was financed through national funds by FCT - Fundação para a Ciência e Tecnologia, I.P. in the framework of the project UID/04564/2025, with DOI identifier 10.54499/UID/04564/2025.


\end{document}